\documentclass[iop]{emulateapj}
\usepackage{graphics,graphicx}
\usepackage{psfig}
\usepackage{times}
\usepackage{comment}
\usepackage{textcomp}
\usepackage{scrextend}
\usepackage{hyperref}

\providecommand{\mb}{$\Delta m_{15}(B)$}

\providecommand{\mu}{$\Delta m_{15}(U)$}

\newcommand{\plotthree}[3]{
  \centering 
  \leavevmode  
  \includegraphics[width=0.3\textwidth]{#1}%
  \hfil
  \includegraphics[width=0.3\textwidth]{#2}%
  \hfil 
  \includegraphics[width=0.3\textwidth]{#3}%
}%

\begin{document}
\title{The Ultraviolet Colors of 
Type I\lowercase{a} Supernovae \\
and their Photospheric Velocities}
\author{Peter~J.~Brown, Jonathan M. Perry, \& Britton A. Beeny}
\affil{George P. and Cynthia Woods Mitchell Institute for Fundamental Physics \& Astronomy, \\
Texas A. \& M. University, Department of Physics and Astronomy, \\
4242 TAMU, College Station, TX 77843, USA }            
\author{Peter A. Milne}
\affil{Steward Observatory, University of Arizona, \\ 933 N. Cherry Avenue, Tucson, AZ 85721, USA}
\author{Xiaofeng Wang}
\affil{Physics Department and Tsinghua Centre for Astrophysics,\\ Tsinghua University, Beijing, 100084, China}

\email{pbrown@physics.tamu.edu}  

\begin{abstract}

We compare ultraviolet (UV) and optical colors of a sample of 29 type Ia supernovae (SNe Ia) observed with the Swift satellite's UltraViolet Optical Telescope (UVOT) with theoretical models of an asymmetric explosion viewed from different angles from Kasen \& Plewa.  This includes mid-UV (1600-2700 \AA; $uvw2$ and $uvm2$) and near-UV (2700-4000 \AA; $uvw1$ and $u$) filters. We find the observed colors to be much redder than the model predictions, and that these offsets are unlikely to be caused by dust reddening.  We confirm previous results that high-velocity SNe Ia have red UV-optical colors.  When correcting the colors for dust reddening by assuming a constant $b-v$ color we find no correlation between the $uvw1-v$ or $u-v$ colors  and the ejecta velocities for 25 SNe Ia with published velocities and/or spectra.  When assuming an optical color-velocity relation, a correlation of 2, and 3.6 $\sigma$ is found for $uvw1-v$ and $u-v$.  However, we find that the correlation is driven by the reddening correction and can be reproduced with random colors which are corrected for reddening.  The significance of a correlation between the UV colors and the velocity is thus dependent on the assumed slope of the optical color-velocity relation. After such a correction, the $uvw1-v$ versus velocity slope is shallower than that predicted by the models and offset to redder colors.  A significant scatter still remains in the $uvw1-v$ colors including a large spread at low velocities.  This demonstrates that the NUV-blue/red spread is not caused by the photospheric velocity.  The $uvm2-uvw1$ colors also show a large dispersion which is uncorrelated with the velocity.

\end{abstract}


\section{Introductions  \label{intro}}

The use of type Ia supernovae (SNe Ia) as cosmological distance indicators relies on our ability to compare their observed brightness to their intrinsic luminosity as standard candles.  This also requires them to be standard crayons, such that their intrinsic colors are known well enough for the effects of dust reddening and extinction to be accurately measured and corrected.  The absolute magnitude is correlated with the light curve shape \citep{Pskovskii_1977,Phillips_1993, Goldhaber_etal_2001} as are the colors  \citep{Riess_etal_1996_mlcs,Phillips_etal_1999}.  Observational evidence points to the amount of $^{56}$Ni as the primary cause of the peak luminosity, light curve shape, and color \citep{Arnett_1982,Stritzinger_etal_2006,Churazov_etal_2014,Scalzo_etal_2014,Diehl_etal_2015}.

While this one-parameter model is effective in constraining distances, understanding the remaining dispersion is of cosmological importance for two reasons.  First, the origin of the dispersion may or may not evolve over the history of the universe, possibly resulting in a bias in distance measures and cosmological parameters.   Second, even if there is no change or bias in the dispersion with redshift, a reduction of the dispersion would increase the precision of measuring distances, necessary to improving constraints on cosmological parameters.

Beyond the light curve shape, SNe Ia can be grouped based on spectroscopic characteristics. Based on the rate of change of the Si II velocity, SNe Ia can be subdivided into high velocity gradient (HVG) or low velocity gradient (LVG) SNe Ia \citep{Benetti_etal_2005}.  HVG SNe Ia tend to have higher  Si II velocities near maximum light, with $\sim$12 $\times 10^6$ m s$^{-1}$ being proposed as a dividing line between high-velocity (HV) and normal SNe Ia\citep{Wang_etal_2009_HV,Wang_etal_2013}.
After \citet{Wang_etal_2009_HV} first noticed that there is a systematic color difference for SNe Ia with different Si II velocities, more evidence has been found supporting intrinsic color differences correlated with the velocities \citep{Maeda_etal_2010,Maeda_etal_2011,Foley_Kasen_2011,Foley_etal_2011_hv,Mandel_etal_2014} though some samples have found the correlation to have low significance \citep{Blondin_etal_2012,Folatelli_etal_2013}.

Asymmetric explosions viewed from different angles has been proposed as a possible cause of this spectral diversity \citep{Kasen_etal_2009,Maeda_etal_2010}.  \citet{Maeda_etal_2011} directly tied asymmetry to the observed nebular line shifts which show correlations with the near-peak velocities and the broad-band colors (see also \citealp{Cartier_etal_2011}).  \citet{Wang_etal_2009_HV} found that using different reddening laws for HVG and normal SNe Ia reduced the dispersion in SN Ia luminosities which could be a signature of different colors or different dust environments.  \citet{Foley_Kasen_2011} suggested the differences could be attributed to intrinsic color differences, with similar reddening laws applying to the low-reddening samples of SNe Ia of both groups.  Asymmetric models from \citet{Kasen_Plewa_2007} were used to show a theoretical connection between velocity and color, and peak velocities were shown to correlate with the peak colors in both the theoretical models and the observations \citep{Foley_Kasen_2011,Foley_etal_2011_hv}.  

However, the fact that HV SNe Ia tend to occur in the inner and luminous regions of their host galaxies indicate that the asymmetric mechanism alone cannot explain the differences in the observed Si II velocity \citep{Wang_etal_2013}.  Instead, the progenitor metallicity, progenitor or companion mass or even explosion models may play important roles.

Despite having peak luminosities with a small dispersion in the optical, SNe Ia show a greater diversity at UV wavelengths \citep{Foley_etal_2008, Ellis_etal_2008,Brown_etal_2010, Wang_etal_2012, Milne_etal_2013, Foley_etal_2016, Brown_etal_2017}.  
The strong effect of many possible progenitor and/or explosion differences (metallicity, density gradient, etc., see e.g. \citealp{Brown_etal_2014}) make UV wavelengths an important regime with which to characterize the diversity of SN Ia explosions.  Understanding the difference amongst the SNe Ia is important for using them as cosmological distance indicators.  The nature of the differences affects whether the mean or scatter in the optical magnitudes evolves with redshift (e.g. due to metallicity differences) or remains constant (e.g. from viewing angle effects).  If the source of the diversity is understood, the magnitude of the observed UV diversity can be compared with models to put bounds on the physical differences.  Towards those goals we are comparing different model expectations and various observed parameters to the growing UV dataset.

In this article we examine axisymmetric models from \citet{Kasen_Plewa_2007} and compare their expectations for magnitudes and velocities with observations of SNe Ia observed with Swift in the UV.  We also search for correlations between the observed photospheric velocity and the UV-optical colors. 
  The paper is outlined as follows:
In Section \ref{section_obs} we briefly describe the sample of Swift SNe Ia used and their velocity data.  
 In Section \ref{section_models} we describe the models used and the relationship between viewing angle and the observable velocities and photometric magnitudes.  The UV photometry is compared with the models in Section \ref{section_modelcomparisons} and the velocities in Section \ref{section_velocitycomparisons}.  We summarize and conclude in Section \ref{section_conclusions}.  

\section{Observations} \label{section_obs}

\subsection{Ultraviolet Photometry}

To examine the observed UV colors, we use SNe Ia photometry from the Swift Optical Ultraviolet Supernova Archive (SOUSA; \citealp{Brown_etal_2014_SOUSA}).  
The Swift UVOT and its photometric calibration are described in \citet{Roming_etal_2005}, \citet{Poole_etal_2008}, and \citet{Breeveld_etal_2010}.  
The filters used in the observations are the mid-UV $uvw2$ and $uvm2$ filters, the near-UV $uvw1$ and $u$ filters, and the optical $b$ and $v$ filters, as shown in the top panel of Figure \ref{fig_filters}.  See \citet{Brown_etal_2010} and \citet{Brown_etal_2016} for detailed discussions of the filter shapes and observed photon distributions for a SN Ia spectrum.  Because the colors have some dependence on the light curve shape \citep{Brown_etal_2010, Wang_etal_2012, Foley_etal_2016}, we restrict our sample to the range of 1.0 $<$ \mb~ $<$ 1.4  and exclude spectroscopically peculiar SNe Ia.  We also restrict the reddening of our sample by including only SNe Ia with $(B-V)_{Bpeak} < $ 0.3 mag to avoid large uncertainties in the reddening correction.  This is the same sample of SNe Ia used in \citet{Brown_etal_2017}.  This Swift SN sample is not unbiased, as the target selection avoided many SNe near the centers of galaxies \citep{Brown_etal_2009} which may result in fewer HV SNe Ia and the majority of nearby SNe in the early years of Swift were found by amateurs and supernova surveys targeting larger galaxies \citep{Leaman_etal_2011}.  Though affecting the relative numbers of HV and normal SNe Ia, we do not believe that this should bias any correlations in the parameters of our sample. We have used this older sample selection to avoid a bias of adding more recent SNe whose spectroscopic data may be available specifically because of a publication bias for extreme or unusual objects. 


\begin{deluxetable*}{lrrrrrrrr}



\tablecaption{Velocities and Colors of Swift SNe Ia}\label{table_velocities}


\tablehead{\colhead{Name} & \colhead{Si II Velocity} &  \colhead{Epoch} &  \colhead{Reference} & \colhead{\mb}  & \colhead{$b-v$} & \colhead{$u-v$} & \colhead{$uvw1-v$} & \colhead{$uvm2-uvw1$} \\ 
\colhead{} & \colhead{(10$^6$ m s$^{-1}$)} & \colhead{(days)} & \colhead{ }  & \colhead{(mag)} & \colhead{(mag)} & \colhead{(mag)} & \colhead{(mag)} & \colhead{(mag)} } 

\startdata

SN2012ht &   9.70 $\pm$   0.10 &   0.00 & \citep{Yamanaka_etal_2014}     &   1.25 $\pm$   0.07 &   0.15 $\pm$   0.04 &  -0.03 $\pm$   0.03 &   1.47 $\pm$   0.04 &   1.89 $\pm$   0.04 \\ 
SN2010gn &   9.91 $\pm$   0.30 &   0.20 & \citep{Maguire_etal_2014}      &   1.24 $\pm$   0.09 &  -0.02 $\pm$   0.05 &  -0.28 $\pm$   0.05 &   1.42 $\pm$   0.07 &    NaN $\pm$    NaN \\ 
SN2005cf &  10.11 $\pm$   0.05 &  -1.20 & \citep{Silverman_etal_2015}    &   1.06 $\pm$   0.02 &   0.06 $\pm$   0.01 &  -0.17 $\pm$   0.07 &   1.51 $\pm$   0.03 &   3.36 $\pm$   0.15 \\ 
SN2011by &  10.27 $\pm$   0.05 &   0.20 & \citep{Silverman_etal_2015}    &   1.11 $\pm$   0.04 &  -0.02 $\pm$   0.02 &  -0.54 $\pm$   0.02 &   1.02 $\pm$   0.02 &   2.60 $\pm$   0.04 \\ 
 SN2015F &  10.30 $\pm$   0.25 &   0.00 & \citep{Foley_etal_2016}        &   1.29 $\pm$   0.18 &   0.23 $\pm$   0.15 &  -0.02 $\pm$   0.15 &   1.68 $\pm$   0.18 &   2.57 $\pm$   0.20 \\ 
SN2011ao &  10.34 $\pm$   0.05 &   0.20 & \citep{Silverman_etal_2015}    &   1.04 $\pm$   0.07 &  -0.00 $\pm$   0.03 &  -0.31 $\pm$   0.03 &   1.51 $\pm$   0.04 &   2.85 $\pm$   0.08 \\ 
SN2005df &  10.42 $\pm$   0.10 &   0.00 & (A.Cikota, in preparation)     &   1.18 $\pm$   0.02 &  -0.07 $\pm$   0.01 &    NaN $\pm$    NaN &   1.48 $\pm$   0.02 &   3.00 $\pm$   0.04 \\ 
SN2007af &  10.56 $\pm$   0.05 &   0.20 & \citep{Silverman_etal_2015}    &   1.20 $\pm$   0.01 &   0.08 $\pm$   0.01 &  -0.13 $\pm$   0.02 &   1.42 $\pm$   0.03 &   2.48 $\pm$   0.14 \\ 
SN2011fe &  10.58 $\pm$   0.05 &  -1.00 & \citep{Silverman_etal_2015}    &   1.05 $\pm$   0.02 &  -0.01 $\pm$   0.01 &    NaN $\pm$   0.01 &   0.99 $\pm$   0.01 &   2.00 $\pm$   0.02 \\ 
SN2011ia &  10.68 $\pm$   0.05 &  -3.00 & \citep{Silverman_etal_2015}    &   1.00 $\pm$   0.09 &  -0.03 $\pm$   0.03 &  -0.78 $\pm$   0.12 &   0.71 $\pm$   0.15 &   2.09 $\pm$   0.15 \\ 
SN2012hr &  10.70 $\pm$   0.10 &   0.00 & \citep{Smartt_etal_2015}       &   1.08 $\pm$   0.07 &   0.00 $\pm$   0.03 &  -0.03 $\pm$   0.03 &   1.64 $\pm$   0.03 &   2.58 $\pm$   0.05 \\ 
SN2008hv &  10.74 $\pm$   0.10 &   1.20 & \citep{Zhao_etal_2015}         &   1.30 $\pm$   0.05 &  -0.07 $\pm$   0.04 &  -0.60 $\pm$   0.04 &   1.05 $\pm$   0.04 &   2.45 $\pm$   0.06 \\ 
SN2008ec &  10.75 $\pm$   0.05 &  -0.20 & \citep{Silverman_etal_2015}    &   1.25 $\pm$   0.03 &   0.21 $\pm$   0.01 &  -0.05 $\pm$   0.03 &   1.65 $\pm$   0.05 &       \nodata           \\ 
SN2013gy &  10.91 $\pm$   0.11 &   1.90 & \citep{Zhao_etal_2015}         &   1.34 $\pm$   0.04 &  -0.04 $\pm$   0.03 &  -0.30 $\pm$   0.03 &   1.45 $\pm$   0.04 &   2.32 $\pm$   0.06 \\ 
SN2006dm &  11.10 $\pm$   0.10 &  -0.65 & HET, this work                 &   1.40 $\pm$   0.05 &   0.09 $\pm$   0.04 &  -0.14 $\pm$   0.04 &   1.54 $\pm$   0.05 &   1.95 $\pm$   0.15 \\ 
 SN2011B &  11.30 $\pm$   0.50 &   0.00 & (X.Wang, in preparation)       &   1.38 $\pm$   0.16 &   0.04 $\pm$   0.10 &  -0.44 $\pm$   0.10 &   1.20 $\pm$   0.08 &   2.32 $\pm$   0.04 \\ 
 SN2008Q &  11.56 $\pm$   0.05 &  -0.29 & \citep{Foley_etal_2011_hv}     &   1.32 $\pm$   0.07 &   0.07 $\pm$   0.03 &  -0.48 $\pm$   0.02 &   0.98 $\pm$   0.03 &   2.06 $\pm$   0.04 \\ 
SN2007co &  11.73 $\pm$   0.05 &  -0.60 & \citep{Silverman_etal_2015}    &   1.09 $\pm$   0.03 &   0.17 $\pm$   0.01 &   0.22 $\pm$   0.03 &   2.05 $\pm$   0.06 &       \nodata           \\ 
SN2006ej &  12.40 $\pm$   0.05 &  -3.70 & \citep{Silverman_etal_2015}    &   1.28 $\pm$   0.04 &   0.10 $\pm$   0.01 &  -0.36 $\pm$   0.14 &   1.11 $\pm$   0.13 &   1.50 $\pm$   0.20 \\ 
SN2013cs &  12.40 $\pm$   0.10 &  -3.00 & \citep{Childress_etal_2016}    &   1.11 $\pm$   0.04 &   0.16 $\pm$   0.03 &   0.19 $\pm$   0.03 &   1.75 $\pm$   0.04 &   2.73 $\pm$   0.08 \\ 
SN2013gs &  12.90 $\pm$   0.20 &   0.00 & (Zhang et al. 2018, ApJ, submitted)       &   1.15 $\pm$   0.06 &   0.03 $\pm$   0.04 &  -0.13 $\pm$   0.04 &   1.61 $\pm$   0.05 &   2.69 $\pm$   0.15 \\ 
SN2010kg &  13.51 $\pm$   0.05 &  -0.10 & \citep{Silverman_etal_2015}    &   1.28 $\pm$   0.07 &   0.28 $\pm$   0.04 &   0.37 $\pm$   0.04 &   1.97 $\pm$   0.05 &       \nodata           \\ 
SN2010ev &  13.60 $\pm$   0.10 &   0.00 & \citep{Gutierrez_etal_2016}    &   1.32 $\pm$   0.07 &   0.30 $\pm$   0.05 &   0.43 $\pm$   0.05 &   2.08 $\pm$   0.06 &   2.46 $\pm$   0.15 \\ 
SN2010gp &  14.00 $\pm$   0.10 &  -3.00 & \citep{Folatelli_etal_2010gp}  &   1.19 $\pm$   0.08 &   0.20 $\pm$   0.04 &   0.37 $\pm$   0.04 &   1.96 $\pm$   0.15 &       \nodata           \\ 
 SN2009Y &  14.88 $\pm$   0.05 &   0.00 & \citep{Folatelli_etal_2013} &   1.04 $\pm$   0.07 &   0.25 $\pm$   0.06 &   0.08 $\pm$   0.02 &   1.73 $\pm$   0.04 &   2.87 $\pm$   0.06 \\ 
SN2013ex &    \nodata          &       \nodata           &             \nodata           &   1.13 $\pm$   0.06 &   0.10 $\pm$   0.03 &  -0.38 $\pm$   0.12 &   1.27 $\pm$   0.15 &   2.25 $\pm$   0.15 \\ 
SN2011im &   \nodata          &       \nodata           &             \nodata           &   1.28 $\pm$   0.08 &   0.29 $\pm$   0.04 &  -0.16 $\pm$   0.14 &   1.76 $\pm$   0.15 &       \nodata           \\ 
SN2009cz &   \nodata          &       \nodata           &              \nodata           &   1.00 $\pm$   0.06 &   0.07 $\pm$   0.04 &  -0.16 $\pm$   0.04 &   1.67 $\pm$   0.05 &   2.84 $\pm$   0.15 \\

\enddata



\end{deluxetable*}

\subsection{Supernova Velocities}

Most Si II $\lambda$ 6355 velocities were collected from the literature including velocities from the classification spectra reported in the Astronomers Telegrams and the International Astronomical Union Circulars and Electronic Telegrams.  The absolute values of the blueshifted velocities are given in units of meters per second (m s$^{-1}$) in Table 1. Some SNe Ia did not have published velocities, but the publicly available spectra from the Open Supernova Archive \citep{Guillochon_etal_2017} or Weizmann Interactive Supernova data REPository (WISEREP; \citealp{Yaron_Gal-Yam_2012}) were used to calculate the velocity based on the absorption minimum of the Si II $\lambda$ 6355 line.
We only used velocities from spectra within 5 days of the time of maximum light in the $B$ band, with phases listed in Table 1.  We note that this range of phases allows a difference of 0.5 to 1 $\times 10^6$ m s$^{-1}$ from the lower to higher velocity SNe Ia \citep{Foley_etal_2011_hv}.  There are 25 SNe Ia in our sample with such velocities.  

One unpublished spectrum of SN~2006dm taken with the Hobby-Eberly Telescope (HET; \citealp{Ramsey_etal_1998} 0.65 days before its maximum was also used.  
This observation used the Low Resolution Spectrograph (LRS; \citealp{Hill_etal_1998}) with a 2\arcsec slit and the GG385 Schott Glass blocking filter.  We used standard IRAF data reduction to bias-subtract, flat field, wavelength calibrate, and flux calibrate the spectrum.  The spectrum will be available via the Weizmann Interactive Supernova data REPository (WISeREP; \citealp{Yaron_Gal-Yam_2012}) and the Open Supernova Catalog (OSC; \citealp{Guillochon_etal_2017}).

\section{Kasen \& Plewa models of an asymmetric explosion} \label{section_models}

\citet{Kasen_Plewa_2007} presented a temporal series of spectra following the radiative transfer of a Chandrasekhar-mass white dwarf exploding as a detonating failed deflagration.  The asymetric explosion results in an egg-shaped ejecta with ejecta velocity structures differing by 6 $\times 10^6$ m s$^{-1}$ from one side to the other.  \citet{Kasen_Plewa_2007} find reasonable matches between the time-series optical photometry and spectroscopy output from their radiative transfer calculations and observations.  They note strong trends of color and ejecta velocity with the viewing angle.  \citet{Foley_Kasen_2011} explicitly show the resulting correlation between color and ejecta velocity
to provide independent, theory-based support for intrinsic color differences in the optical between low and high-velocity SNe Ia.  In this article we look for similar correlations with UV colors predicted by \citet{Foley_Kasen_2011} based on the optical properties studied from the \citet{Kasen_Plewa_2007} models.  

\begin{figure} 
\resizebox{8.8cm}{!}{\includegraphics{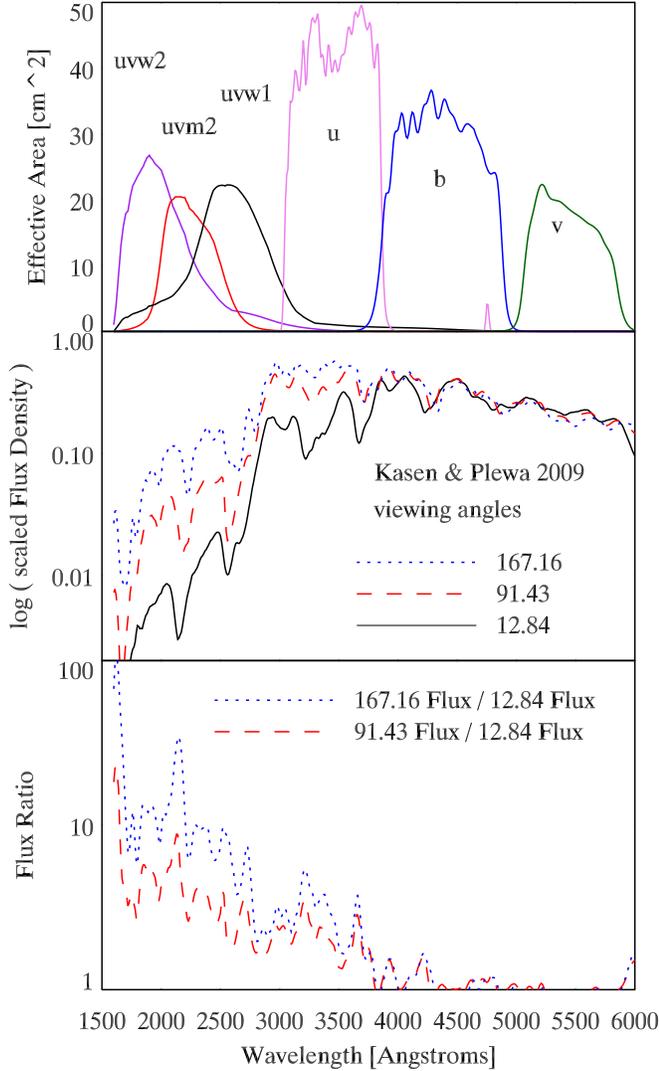}   }
\caption[Results]
        { Top Panel: Swift UVOT filter effective area curves.
		Middle Panel: Theoretical spectra from \citet{Kasen_Plewa_2007} from three different viewing angles.
		Bottom Panel: Flux ratios made from dividing two of the above spectra from the lowest flux spectrum.
 } \label{fig_filters}    
\end{figure}
\begin{figure} 
\resizebox{8.8cm}{!}{\includegraphics{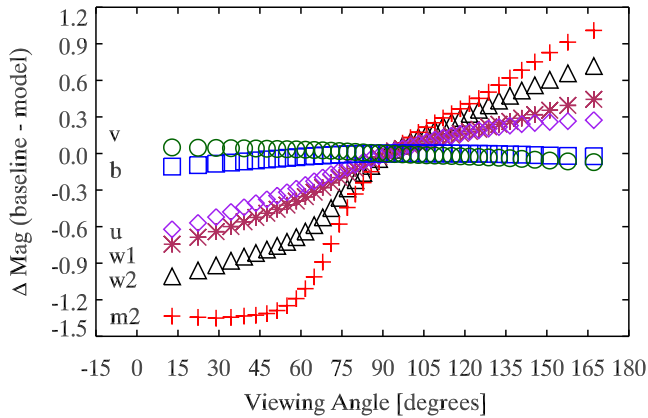}   }
\caption[Results]
        {Magnitude differences in the UVOT filters for the \citet{Kasen_Plewa_2007} model from spectra simulated for different viewing angles.
 } \label{fig_modelmags}    
\end{figure}

To visualize how the model spectra correspond to the photometric observations, Figure \ref{fig_filters} shows the UVOT filters, synthetic spectra from \citet{Kasen_Plewa_2007} for three different viewing angles, and flux ratios made by dividing the other two example spectra by the lowest angle (and lowest flux) spectrum.  The change in viewing angle has limited effect in the optical but the differences generally grow as one considers shorter wavelengths.  To compute the magnitude differences between the models, we performed spectrophotometry on the model spectra using the UVOT filter curves from \citet{Breeveld_etal_2011}.  The magnitude of the 90\textdegree{} model is subtracted from each to highlight the differences in Figure \ref{fig_modelmags}.  The smallest viewing angles have the reddest colors.  The $B-V$ difference is small, but the Swift $u$ band and the shorter wavelength filters are strongly affected.  The Swift $uvm2$ exhibits the largest effect, greater than 2.5 mag.  The $uvw2$ filter has a shorter central wavelength, but with an optical tail which is significant for red objects \citep{Brown_etal_2010,Brown_etal_2016}.  Because of the optical tail or ``red leak'', the $uvw2$ is not as sensitive as the $uvm2$ filter to UV spectral changes, though Figure \ref{fig_modelmags} demonstrates it would still be effective on its own in identifying extreme UV behavior.

\begin{figure}[h]
\resizebox{8.8cm}{!}{\includegraphics{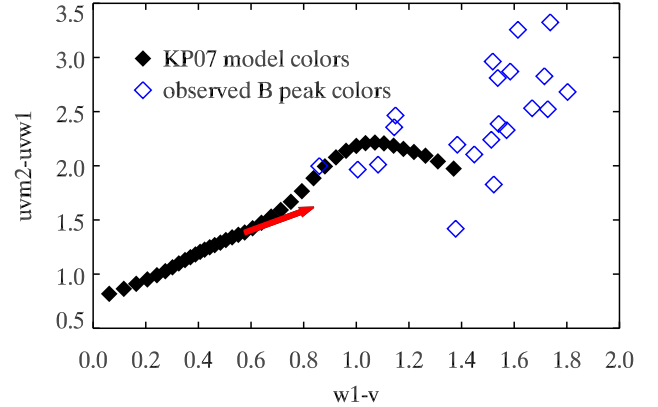}   }
\caption[Results]
        {  The colors from the models of \citet{Kasen_Plewa_2007} are compared to the observed peak colors of our sample.  The red arrow indicates the direction resulting from MW reddening with R$_V$=3.1 \citep{Cardelli_etal_1989}.  To be consistent, the whole sample would have to be reddened by about an E(B-V) color excess of 0.2-0.3 mag, inconsistent with the optical colors.
 } \label{fig_modelexcolors}    
\end{figure}

\section{Comparisons between the Kasen \& Plewa model and UV observations}\label{section_modelcomparisons} 

An accurate analysis of UV data requires a proper treatment of extinction.  
Determining the correct intrinsic colors requires a correction for the right amount of dust and the appropriate wavelength dependence of the extinction, i.e. the reddening law.  This is done in the optical by assuming a set of SNe Ia to be unreddened; comparing reddened and unreddened SNe Ia yields a color excess in different bands.  
This can also be done in the UV for individual SNe (e.g. SN~2014J; \citealp{Amanullah_etal_2014,Foley_etal_2014,Brown_etal_2014J}) or for larger samples \citep{Amanullah_etal_2015}.  However, the intrinsic diversity of SNe Ia in the UV \citep{Brown_etal_2010, Wang_etal_2012, Milne_etal_2013, Foley_etal_2016, Brown_etal_2017} and the uncertain extrapolation of extinction laws into the UV makes an accurate correction complicated.  To study the intrinsic diversity, one must correct for extinction.  To correct for extinction, one must understand the intrinsic colors and their diversity.  This circular regression is not easily resolved.

In this work, we will attempt to display the data in raw and corrected form such that we can at least qualitatively assess the strength of any relationship between the UV colors and the ejecta velocities.  Searching for a relationship between the intrinsic UV color of SNe Ia and their ejecta velocities is further complicated by the possibility of a relationship between the optical colors and the velocities.  Potential scenarios include the assumption of a relationship which does not exist, a relationship which is ignored but does exist, or a relationship which exists but which is corrected for inaccurately.  In any of these cases, the extrapolation of the correction into the UV (or lack of correction) could result in a biased interpretation of the data.  

Figure \ref{fig_modelexcolors} displays the observed $uvm2-uvw1$ and $uvw1-v$ colors of the \citet{Kasen_Plewa_2007} models for all viewing angles with filled symbols.  Also shown is a reddening vector corresponding to a Milky Way-like extinction law \citep{Cardelli_etal_1989} with R$_V$=3.1 and $E(B-V)=0.1$.  Interestingly, the colors of the lower left points, which would correspond to a lower viewing angle and smaller ejecta velocities, seem to follow the reddening vector.  This applies only to the colors, however, and would result in an incorrect color excess and distance measurement, as the $v$-band data does not get fainter as it would with dust extinction but actually increases in brightness in these models, as shown in Figure \ref{fig_modelmags}.  A combination of dust reddening/extinction and velocitiy reddening with no dimming in the redder bands would lead to a lower R$_V$ value as observed \citep{Kessler_etal_2009,Burns_etal_2014}.  The redder bump in the $uvm2-uvw1$ colors could be misinterpreted as a strengthening of the 2175 ~\AA~bump common to MW sight lines.  However, these colors are all intrinsic to the SN model itself.  These possible effects highlight the need to disentangle intrinsic color differences from dust reddening \citep{Chotard_etal_2011, Scolnic_etal_2014,Sasdelli_etal_2016,Mandel_etal_2017}.

The open diamonds in Figure \ref{fig_modelexcolors} show the observed colors of our Swift SN Ia sample.  If they represented observations of SNe Ia following the \citet{Kasen_Plewa_2007} model they would have similar colors with the reddened SNe Ia pushed up and to the right in the direction of the reddening vector.  The observed colors, however, are all offset to the red in both colors, and the $uvm2-uvw1$ colors in particular appear to have a much larger scatter.

\section{Comparisons between the Si II Velocities and UV colors}\label{section_velocitycomparisons} 

\begin{figure*} 
\plotthree{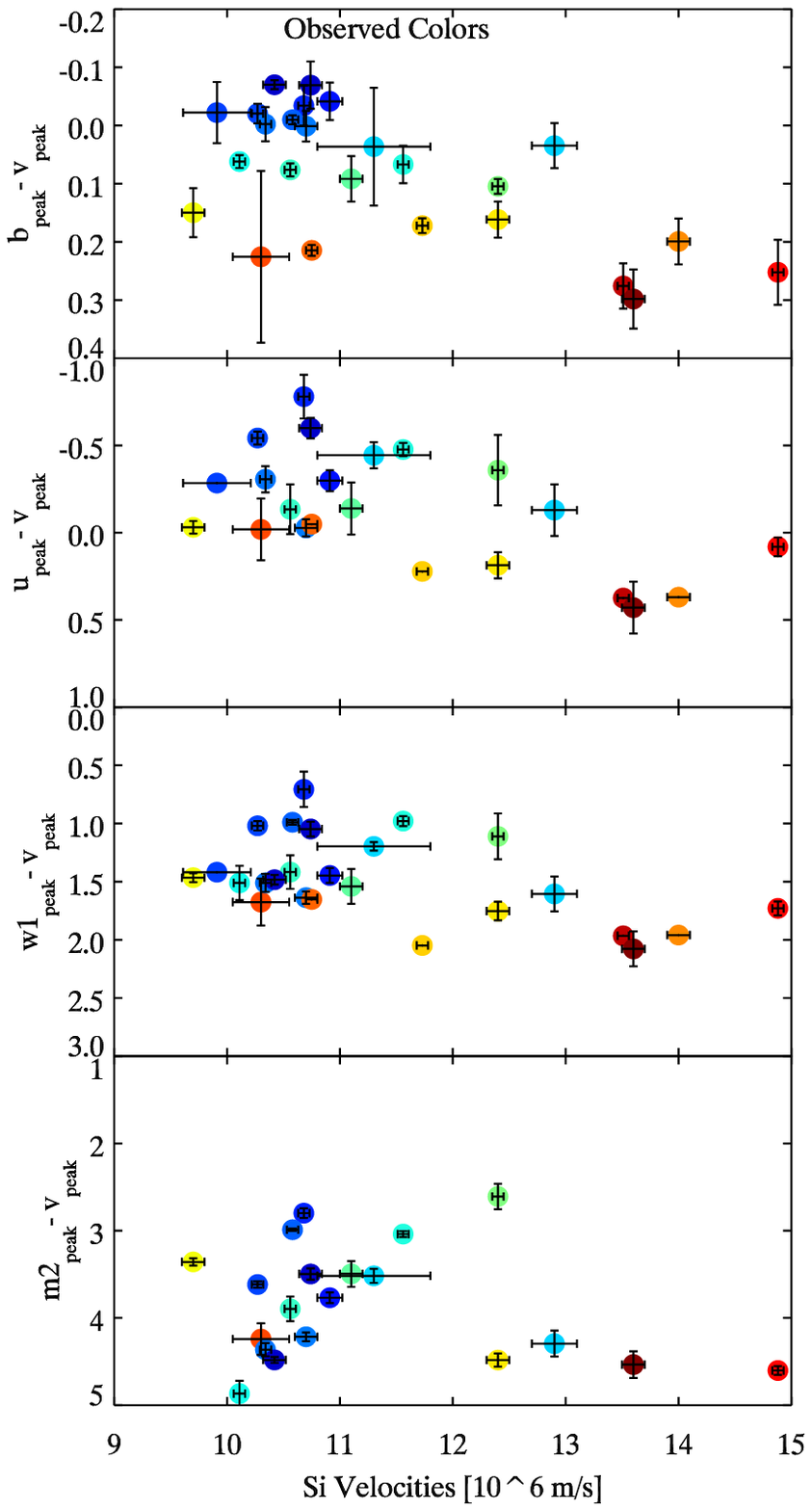}{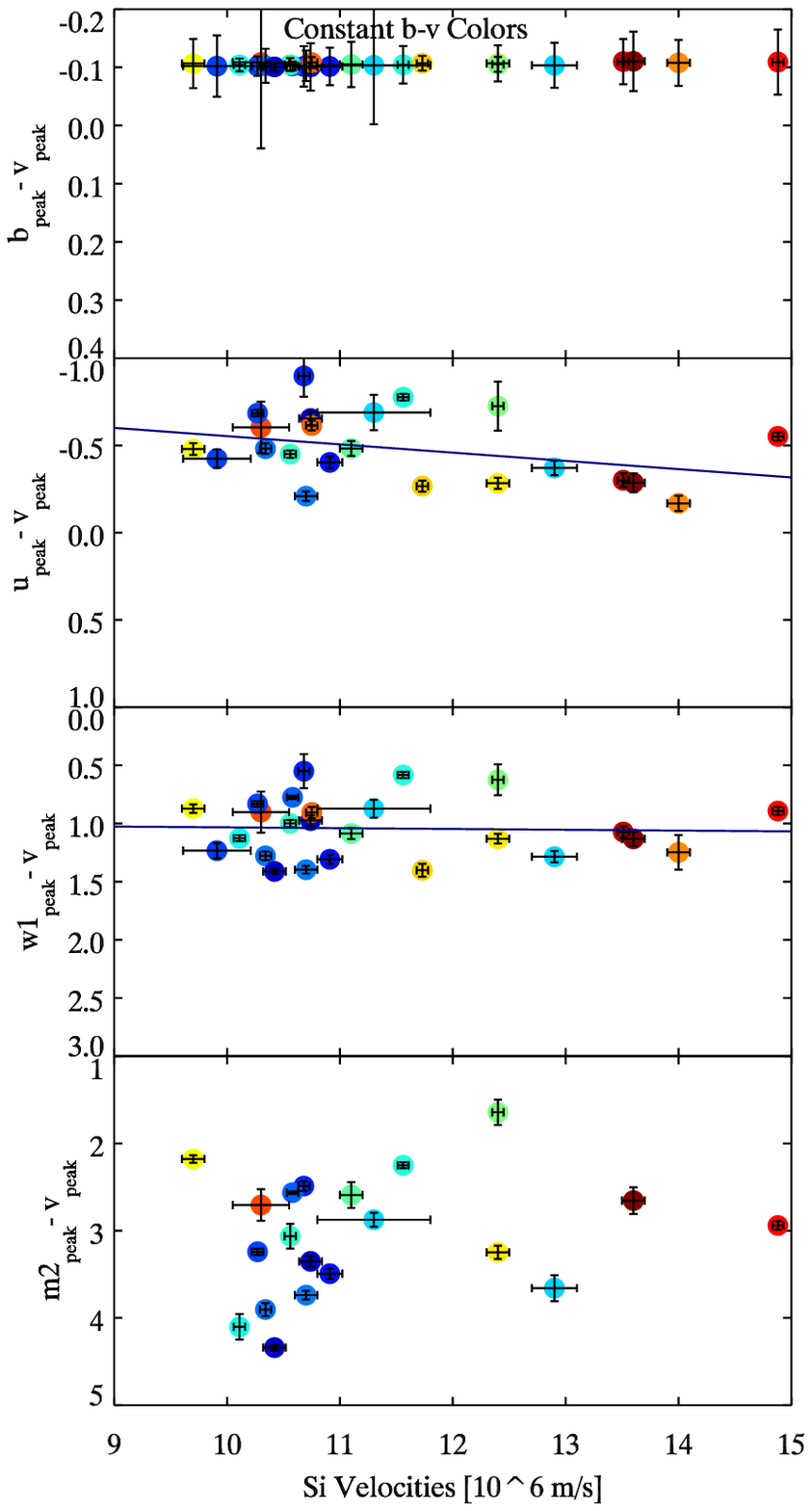}{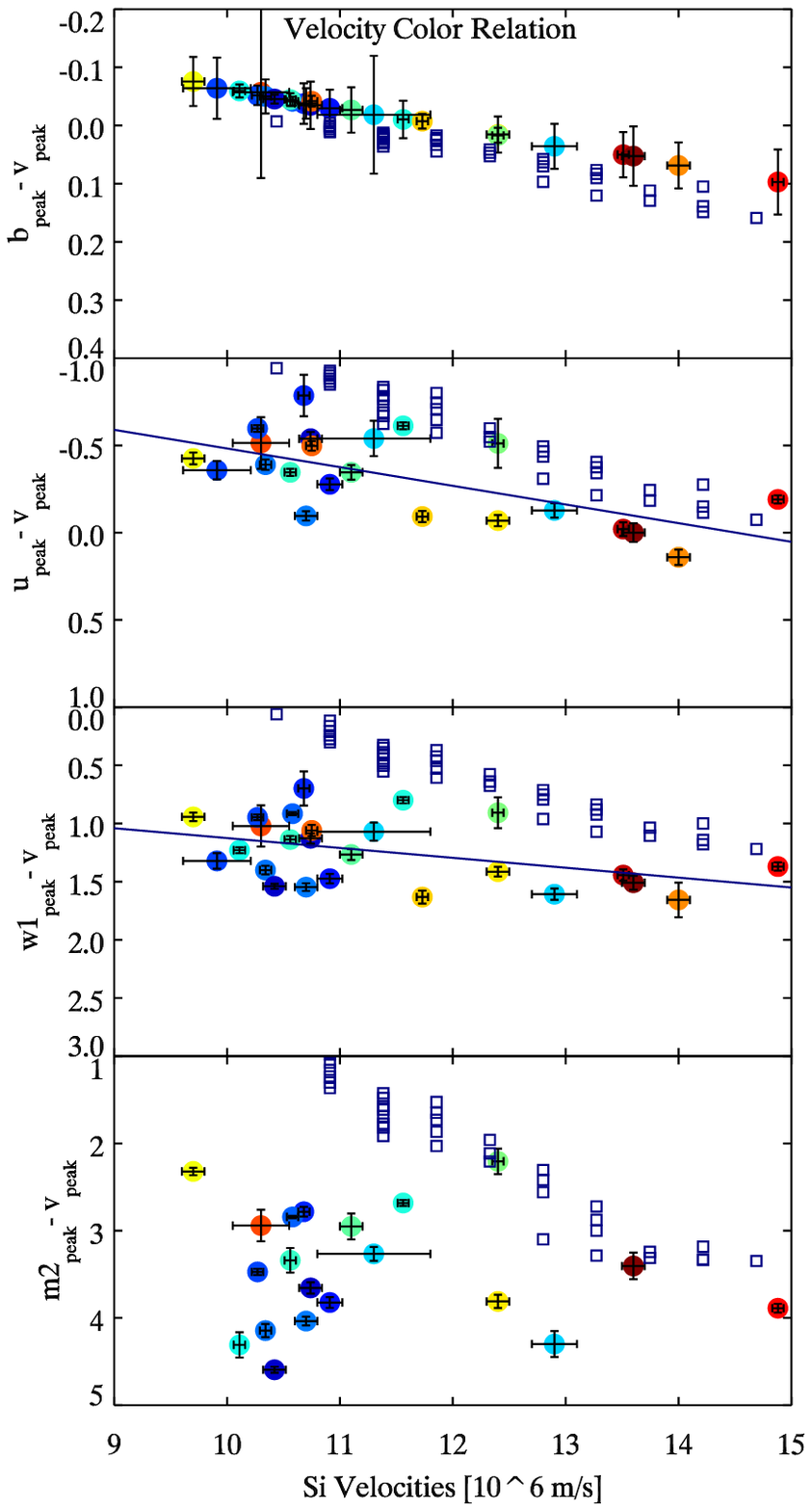} 
\caption[Results]
        {Left: The observed $b-v$, $u-v$, $uvw1-v$, and $uvm2-v$ colors interpolated at the time of maximum light in the $B$ band are plotted with respect to the velocity of Si II $\lambda$6355 within 5 days of maximum light.  The symbols are colored based on their observed $b-v$ color; this color scheme is preserved in the other panels to represent how large of a reddening correction was made.  Middle: The same colors are corrected for reddening by assuming a single $b-v$ color and a Milky Way-like extinction law with R$_V$=3.1.  Solid lines show the median slope determined by our MCMC regression fitting.
Right: The same colors are corrected for reddening by assuming a relationship between the $b-v$ color and the Si II velocity from equation 11 of \citet{Foley_etal_2011_hv}.  Open squares show the colors of the \citep{Kasen_Plewa_2007} models near maximum light.
 } \label{fig_colorvelocities}    
\end{figure*}

Recognizing that there may be an offset in the absolute colors between the models and the observations, as \citet{Foley_Kasen_2011} found for the $B-V$ colors, we now proceed to examine the possibility of a relationship between the colors and ejecta velocity, as predicted by the model and observed by Swift.  
In Figure \ref{fig_colorvelocities} we plot the expansion velocities and the optical and pseudocolors from the peak magnitudes (with respect to $v$-band peak magnitude) with different treatments for extinction.  For the Si II $\lambda$6355 velocities we use the absolute value of the velocity from spectra taken within 5 days (before or after) of maximum light in the B band.  The left panel shows colors uncorrected for extinction.  
The middle panel shows the colors if a constant eak $b-v$ pseudocolor is assumed and any $b-v$ color excess corrected for using a MW extinction law \citep{Cardelli_etal_1989} with R$_V$=3.1.  
The right panel shows the colors corrected for reddening by assuming a relationship between the $b-v$ color and the Si II velocity from equation (11) of \citet{Foley_etal_2011_hv} and extinction coefficients derived for a UV-optical spectrum of SN~2011fe near peak \citep{Mazzali_etal_2014} and a MW extinction law with R$_V$=3.1 \citep{Cardelli_etal_1989}. 
Because we are examining colors rather than absolute magnitudes, a choice of R$_V$=3.1 rather than 1.7 does not have a large effect on the results even if a different extinction law might be more appropriate for HV SNe \citep{Wang_etal_2009_HV}.

In the left panel, the observed colors confirm a finding by \citet{Milne_etal_2013}; namely, all of the high velocity (v$>12$ Mm s$^{-1}$) SNe Ia are NUV-red ($w1-b > 0.8$ at peak) when uncorrected for reddening.   SNe Ia with red observed colors of  $B-V$ $>$ 0.3 have already been removed.  A few higher-velocity SNe Ia are right near the $B-V$=0.3 mag cut-off with few with low $B-V$ values.  The paucity of higher-velocity SNe Ia with blue $B-V$ colors could be intrinsic \citep{Wang_etal_2009_HV, Foley_Kasen_2011} or due to reddening in their environment \citep{Wang_etal_2013}.  The higher-velocity SNe Ia have redder NUV-optical colors than the normal SNe Ia, suggesting there could be a trend of the NUV-optical colors with velocity.  However, at the low velocity end there is still quite a spread in the colors, suggesting that the velocities are not strongly correlated with the NUV-red/blue distinction \citep{Milne_etal_2013,Brown_etal_2017}.  In other words, low velocity SNe Ia can be either NUV-red or NUV-blue.  

The lack of NUV-blue SNe Ia with higher velocities could be a physical distinction or it could be that there is a reddening effect correlated with velocity which effects both groups.  A reddening effect of velocity on colors could make an otherwise NUV-blue SN Ia appear NUV-red similar to the effect of dust reddening \citep{Brown_etal_2017}.
Just as one might need to correct the color of a SN Ia for reddening before deciding whether it is NUV-red or NUV-blue, one might also need to correct for velocity effects.  If there is not a separation of two groups but merely a continuum of NUV-optical colors, a similar reasoning still applies.  One needs to correct a SN Ia for reddening and velocity effects to determine where it lies in a continuum of NUV-optical colors determined by an additional parameter beyond light curve shape, dust reddening, and velocity.

Fits of color versus velocity were calculated using linear regression with errors in the velocity and magnitudes with 10,000 Markov Chain Monte Carlo (MCMC) tries using an IDL routine LINMIX\_ERR from \citet{Kelly_2007}.  Table 2 contains the weighted mean ($\pm$ one standard deviation or $\sigma$) before the fitting, the mean intrinsic scatter ($\pm$ 1 $\sigma$) from the MCMC fits, and the mean slope ($\pm$ 1 $\sigma$) from the MCMC fits for each of the four colors and three sets of corrections (or lack thereof) for reddening.  Most relevant to this work is a comparison between the standard deviation of the mean without the fit and the mean intrinsic scatter of the fit, the change in the intrinsic scatter with the different reddening corrections, and the slope.   

After correcting for reddening by assuming a constant $b-v$ color, the $uvw1-v$ and $u-v$ colors have no significance slopes of redder colors with increasing velocity, 0 and 1.6 $\sigma$, respectively, when considering the fraction of posterior pulls with negative slopes.  The slope and significance increases if an optical color-velocity trend is assumed, with $uvw1-v$ and  $u-v$ yielding negative slopes with significances of 2. and 3.6 $\sigma$, respectively.  The change in slope is not unexpected as redder objects with higher velocities are corrected less in their UV magnitudes when an optical color correlation is assumed.  

The strong dependence of the slope on the reddening correction was
confirmed by shuffling the SN velocities before the reddening correction and fitting 100 times.  The mean slope for $uvw1-v$ was 0.071 with a standard deviation of 0.08 in the MCMC trials (compared to 0.08 $\pm$ 0.04 for the velocity-corrected colors), with only three trials having a negative slope.  This apparent $\sim$2 $\sigma$ relationship shows a UV correlation could be created where none exists just by the assumption of an optical correlation.   This highlights the need for a clear understanding of the optical colors, since the UV colors depend so much on them.  

The scatter in the colors does not change significantly when comparing the standard deviation of the colors after correction to a constant color without and with a color-velocity relation.   The intrinsic scatter determined from the fitting in $uvm2-v$, $uvw1-v$, and $u-v$ is 0.56, 0.07, and 0.04 mag.
The lack of change is also not unexpected, as the reddening correction merely increases the slope of the velocity-color relation, and the points scatter about it the same.  The constant intrinsic scatter actually drives the apparent increase in the significance of the slopes when going from a constant color correction to a velocity-optical color correction. The scatter of the MCMC-determined slopes remains the same, but the value of the slope has been changed.

Just as a weak correlation of the UV colors with velocity could be artificially created by assuming a steeper optical correlation, a correlation can be weakened by assuming a shallower than actual optical correlation.  If such a correlation in the $b-v$ colors is implicitly removed, as we did in the middle panels of Figure \ref{fig_colorvelocities}, the UV-optical colors of higher-velocity SNe Ia are overcorrected for reddening.  The effect can be especially difficult to untangle if the wavelength dependence of the intrinsic variations is similar to a dust reddening law. As shown in Figure \ref{fig_modelexcolors}, the color differences (in effect a broad-band reddening law) resulting from different viewing angles has the same color-color slope as dust reddening.  Some models for SNe Ia with changing metallicity also show similar behavior as dust reddening laws (e.g. Figure 10 in \citealp{Brown_etal_2015}), complicating the separation of intrinsic colors from dust reddening.  Misinterpretation of the intrinsic optical color could lead to erroneous extinction corrections and ultimately wrong luminosity distances because for the viewing angle differences cause the bluer optical colors to be affected but not the optical luminosity in redder bands.

Plotted as open squares in the right panel of Figure \ref{fig_colorvelocities} are the predicted velocities and colors from \citet{Kasen_Plewa_2007}.  Correcting the $b-v$ colors according to \citet{Foley_etal_2011_hv} gives them approximately the same slope as the model colors versus velocity with the observed colors offset to the blue by $\sim0.05$ mag.  
The observed $uvw1-v$ colors have a shallower slope with respect to velocity than the models and are offset to the red by $\sim1$ mag.  This offset is much larger and in the opposite direction from that resulting from overcorrecting the reddening compared to the model colors.  The observed $uvm2-uvw1$ colors are also redder than the models by $\sim$2 mag.  They do not show the steep slope with velocity predicted by the \citet{Kasen_Plewa_2007} models, though the number of higher velocity Swift SNe Ia detected in the $uvm2$ filter is lower.  The offset in absolute colors would suggest caution before arriving at strong conclusions based on trends in the model UV colors, as the models have deficiencies in the UV.


\section{Summary and Conclusions} \label{section_conclusions}

We have taken a set of asymmetric SN Ia explosion models \citep{Kasen_Plewa_2007} to determine the predicted effects on the magnitudes and colors in the UV and optical bands of Swift UVOT.   The flux of the models is higher in the UV than observed, with correspondingly bluer colors.  The models do predict the signature of asymmetry to be stronger at shorter wavelengths, well-probed by UVOT, but with color trends similar to dust reddening.  

We also compare the observed UV colors with the near-peak Si II velocities.  We do find a tendency for higher velocity SNe Ia to have redder observed UV colors  as first noted by \citet{Milne_etal_2013}.  However, we find that correcting for dust reddening dominates over an intrinsic correlation with velocity.  In other words, the assumed correlation between the optical colors and velocities determines the strength of a correlation between the UV colors and the velocities.  Thus we cannot conclude whether such a correlation exists.  We do see a spread in UV-optical colors for low/normal velocity, indicating that the UV-optical dispersion is not related to the photospheric velocity as measured from the Si II line.

We encourage color-independent means of constraining the dust reddening, e.g. from polarization, Na I D absorption (\citealp{Munari_etal_1997,Poznanski_etal_2012}; but see also \citealp{Poznanski_etal_2011} for the impact of low resolution spectra and \citealp{Phillips_etal_2013} for cases of SNe Ia with higher Na ID absorption than predicted by reddening), and from the diffuse interstellar bands \citep{Phillips_etal_2013}.  
Color-independent means of inferring the UV colors would also better allow us to determine whether HV SNe Ia could be physically related to the NUV-blue SNe Ia with low velocities but with the velocities causing the colors to be redder similar to reddening.  Unfortunately the observed trend for NUV-blue SNe Ia to have detections of CII $\lambda6580$ \citep{Thomas_etal_2011,Milne_etal_2010} is harder to test for HV SNe because of doppler broadening of the Si II line \citep{Parrent_etal_2011,Folatelli_etal_2012}.

We have studied possible correlations of the UV colors with the Si velocity, due to the potential color differences found previously in the optical.  Stronger correlations might be found with the bluer Si and Ca H\&K lines, due to the strength of the Si/Ca absorption within the Swift u and uvw1 bands, or the strength of the high velocity features or high velocity wings of the photospheric absorption from the outer region where the UV photons escape. The UV flux in the Swift mid-UV filters is especially sensitive to the density gradients in the very outer laters, extending at least 9 days after maximum light \citep{Sauer_etal_2008,Brown_etal_2014}.  Similarly, high velocity features may also tell us about the regions of the ejecta from which the UV light originates.
 \citet{Wang_etal_2012} found that artificially increasing the velocity of the W7 model actually resulted in an increase in the UV flux--the opposite effect seen here--because the UV emission region had a larger radius.  Thus velocity cannot be considered in isolation as a cause, but only in the context of a physical model resulting in a change in the observed velocities and the associated changes in density structure and composition associated with the physical model changes.  A larger sample and further study of UV spectra (continuing from \citealp{Foley_etal_2008_UV,Bufano_etal_2009,Smitka_2016,Foley_etal_2016, Pan_etal_2018}) would be of great benefit in differentiating velocity and other variables which may or may not be linked, such as metallicity, density structure, explosion energy, and progenitor system characteristics.

We have used the \citet{Kasen_Plewa_2007} models which have asymmetry as the origin of the velocity differences.   The failure of the models and the behavior in the UV could be due to a number of reasons: The physical scenario represented in the model might not be that realized in nature.  The physical scenario might be correct but the model fails to accurately reproduce the observational characteristics due to incomplete line lists, missing physics, or different initial conditions unrelated to the parameters being studied.  The physical scenario might be correct but of a different magnitude than probed.  
It is not clear which is the cause for these models, though we note that first principle models have difficulty matching the observed UV flux of SNe Ia \citep{Lentz_etal_2000,Kasen_2010} and the models tuned to match observations \citep{Sauer_etal_2008,Walker_etal_2012,Mazzali_etal_2014} must typically use a different density profile and have metallicity and other free parameters.  
With those difficulties and degeneracies between the multiple effects in the UV \citep{Brown_etal_2014} it is hard to rule out physical scenarios or explosion models based on the inability to match UV observations and judge the applicability of models at other wavelengths. The \citet{Kasen_Plewa_2007} models also predict a larger than observed polarization \citep{Wang_Wheeler_2008}, so asymmetry may be only one component affecting the velocity differences.  
\citet{Wang_etal_2013} find HV SNe Ia to be more common near the cores of larger, more massive galaxies, indicating a difference in progenitors rather than just viewing angle.  These progenitors could be more metal rich, consistent with various arguments made by \citet{Pignata_etal_2008} for a metallicity origin of the differences before 
favoring the more extended explosive-burning front advocated by \citet{Benetti_etal_2004}. 
\citet{Foley_etal_2012_prog} found that HV SNe Ia tend to have blueshifted NaID absorption, also suggestive of a difference in the progenitor and/or its environment.
Resolving the difference is important because observed differences of asymmetric explosions viewed at different angles will look similarly different throughout the history of the universe.  Metallicity, however, will systematically change through the history of the universe, resulting in a bias in the mean magnitudes of SNe Ia and a change in the scatter with redshift.

The Swift SN Ia sample is not unbiased, as some SN searches, particularly those in the early years of Swift, targeted large galaxies and missed SNe from faint, low mass hosts now seen in untargeted surveys. Additionally, some surveys are less sensitive to finding SNe near the cores of galaxies and early Swift SN trigger criteria recommended a separation of 10\arcsec from nearby stars and the host galaxy nucleus to avoid issues with coincidence loss \citep{Brown_etal_2009}.  As HV SNe Ia have been found to be preferentially located near the centers of their host galaxies \citep{Wang_etal_2013}, this would reduce the number of HV SNe being discovered and further reduce the number being observed by Swift.  The All-Sky Automated Survey for SuperNovae (ASAS-SN; \citealp{Shappee_etal_2014} is more efficient than other amateur and professional searches at finding SNe close to the host nucleus \citep{Holoien_etal_2017}.  Between Fall 2016 and Spring 2018 a program was  to observe with Swift/UVOT a volume-limited sample of SNe Ia within z$<$0.02 without regard to host-SN separation.  Hardware-window modes \citep{Roming_etal_2005,Poole_etal_2008} are used to mitigate the effects of coincidence loss as in \citet{Brown_etal_2012_11fe,Brown_etal_2014J}.  This program will reduce any bias in Swift SN Ia properties and should increase the numbers of higher-velocity SNe Ia available for future studies.



\acknowledgements

The Swift Optical/Ultraviolet Supernova Archive (SOUSA) is supported by NASA's Astrophysics Data Analysis Program through grant NNX13AF35G.
This work made use of public data in the {\it Swift} data
archive and the NASA/IPAC Extragalactic Database (NED), which is
operated by the Jet Propulsion Laboratory, California Institute of
Technology, under contract with NASA.  
X. Wang is supported by the National Natural Science Foundation of China 
(NSFC grants 11325313 and 11633002), 
and and the National Program on Key Research and Development Project (grant no. 2016YFA0400803)



\bibliographystyle{apj}


\clearpage

\begin{deluxetable*}{llrrr}
\tablecaption{Results of Color-Velocity Regression Fitting}\label{table_fits}

\tablehead{\colhead{Color} & \colhead{Method} &  \colhead{Weighted Mean\footnote{\label{a}The mean and standard deviation of the color for all of the data, ignoring any fit relation.  It is the standard deviation which is usefully compared to the scatter after fitting.}} &  \colhead{Fit Scatter} & \colhead{Slope}  \\ 
\colhead{} & \colhead{} & \colhead{(mag)} & \colhead{(mag)} & \colhead{(mag Mm$^{-1}$ s)} } 

\startdata

$b-v$ & observed &   0.05 $\pm$   0.11 &   0.01 $\pm$   0.00 &   0.05 $\pm$   0.01 \\
$b-v$ & dereddened &  -0.10 $\pm$   0.00 &   0\footnote{\label{b}The $b-v$ slope are set in the reddening corrections, so these values are defined in the fit and just given for reference.} &  0\footref{b} \\
$b-v$ & velocity-corrected &  -0.03 $\pm$   0.05 &   0.00\footref{b} &   0.03\footref{b} \\
 
$u-v$ & observed &  -0.13 $\pm$   0.33 &   0.08 $\pm$   0.04 &   0.14 $\pm$   0.04 \\
$u-v$ & dereddened &  -0.38 $\pm$   0.20 &   0.04 $\pm$   0.02 &   0.05 $\pm$   0.03 \\
$u-v$ & velocity-corrected &  -0.26 $\pm$   0.24 &   0.04 $\pm$   0.02 &   0.11 $\pm$   0.03 \\

$uvw1-v$ & observed &   1.33 $\pm$   0.36 &   0.10 $\pm$   0.04 &   0.13 $\pm$   0.05 \\
$uvw1-v$ & dereddened &   1.00 $\pm$   0.25 &   0.07 $\pm$   0.03 &   0.00 $\pm$   0.04 \\
$uvw1-v$ & velocity-corrected &   1.17 $\pm$   0.28 &   0.07 $\pm$   0.03 &   0.08 $\pm$   0.04 \\
 
$uvm2-v$ & observed &   3.41 $\pm$   0.66 &   0.49 $\pm$   0.21 &   0.12 $\pm$   0.13 \\
$uvm2-v$ & dereddened &   2.86 $\pm$   0.69 &   0.57 $\pm$   0.26 &  -0.11 $\pm$   0.14 \\
$uvm2-v$ & velocity-corrected &   3.18 $\pm$   0.68 &   0.56 $\pm$   0.26 &   0.05 $\pm$   0.14 \\

\enddata



\end{deluxetable*}

\end{document}